# OCCIA LAB

*To discover the causes of social, economic and technological change*

**CocciaLab Working Paper 2018 – No. 32**

# How do public research labs use funding for research? A case study

Mario COCCIA

CNR -- NATIONAL RESEARCH COUNCIL OF ITALY

&

ARIZONA STATE UNIVERSITY

# How do public research labs use funding for research? A case study

*Mario Coccia*[1]

CNR -- NATIONAL RESEARCH COUNCIL OF ITALY &
ARIZONA STATE UNIVERSITY

*E*-mail: mario.coccia@cnr.it

Current Address: COCCIA*LAB* at CNR -- NATIONAL RESEARCH COUNCIL OF ITALY
Collegio Carlo Alberto, Via Real Collegio, 30, 10024-Moncalieri (Torino), Italy

Mario Coccia : http://orcid.org/0000-0003-1957-6731

This paper discusses how public research organizations consume funding for research, applying a new approach based on economic metabolism of research labs, in a broad analogy with biology. This approach is applied to a case study in Europe represented by one of the biggest European public research organizations, the National Research council of Italy. Results suggest that funding for research (state subsidy and public contracts) of this public research organization is mainly consumed for the cost of personnel. In addition, the analysis shows a disproportionate growth of the cost of personnel in public research labs in comparison with total revenue from government. In the presence of shrinking public research lab budgets, this organizational behavior generates inefficiencies and stress. R&D management and public policy implications are suggested for improving economic performance of public research organizations in turbulent markets.

---

[1] I gratefully acknowledge financial support from the CNR - National Research Council of Italy for my visiting at Arizona State University (Grant CNR - NEH Memorandum n. 0072373-2014 and n. 0003005-2016) where this research started in 2016.

CONTENTS

Coccia M. (2018) How do public research labs use funding for research? A case study

*CocciaLab Working Paper 2018 – No. 32*

# Introduction

This paper has two goals. The first is to analyze *how* public research organizations use funding for research. The second is to suggest best practices of R&D management in public research organizations to sustain scientific production. These topics are basic in the field of management because Public Research Organizations (PROs) produce scientific research that can support long-run competitive advantage of countries (Crow and Bozeman, 1991, 1998; Piccaluga and Chiesa, 2000). PROs are supported by governments with public funds and unless properly financed and governed are doomed to organizational inefficiencies and low scientific production (Coccia, 2012; Coccia *et al.,* 2015). Current economic crisis in many Western economies is generating a high government debt between nations. Countries, to cope with this problematic situation, tend to apply policies of fiscal compact that triggers automatic cuts in spending in the absence of deficit reductions (Barseghyan and Battaglini, 2016). One of the consequences of these economic policies is cuts to public research sector, reduction of investments in R&D and in the budgets of PROs (Boden et al., 2004; Cruz-Castro and Sanz-Menéndez, 2016, 2018; Sanz-Menéndez and Van Ryzin, 2015; Coccia, 2012; Coccia, 2017; Coccia *et al.,* 2015). In this context, a cost analysis of public research organizations, in the presence of reduced state subsidies and public contracts, is more and more important for supporting scientific research production, operations and survival of public research labs in basic and applied research fields. The literature has analyzed several aspects of public research organizations (*cf.,* Coccia 2012; Crow and Bozeman, 1987, 1991, 1998). However, it is hardly known *how* public research organizations use funding for research in the presence of shrinking public research lab budgets and rigid organizational structures.

In light of the continuing importance of the studies of R&D management on public research organizations (Coccia, 2001, 2004, 2012), this study focuses specifically on following research questions:

☐ How do public research organizations use public funding to produce scientific outputs?

- How do trends of public funding and operational costs evolve within public research organizations? Which factors matter most?

The underlying problem of these research questions is to analyze the consumption of public research funding within public research bodies. This study suggests, in a broad analogy with biology, an approach called economic metabolism of labs to analyze how they use and/or transform inputs (i.e., public funding for research, equipment, activity of human resources) to produce scientific research, innovative outputs and technology and innovation services (cf., Readman et al., 2018, *passim*)[2]. Results of this study can detect organizational problems of PROs for designing a fruitful R&D management directed to control and improve allocation of public research funding within PROs. In order to position the study here into the literature, next section presents the theoretical framework of this paper.

**Theoretical background**

Research laboratories and organizations can be considered complex systems of economic and human resources (inputs), and production processes directed to generate scientific research, technology transfer and discoveries (outputs)[3]. Coccia (2001, p. 454) argues that research organizations are a combination of elements and human resources are the most important element that creates knowledge, which is originated at individual level and/or groups and spread within and outside organizational systems (cf., Nonaka, 1994). In this context, the concept of metabolism, in a broad analogy, can be applied from biology to management to analyze the processes running from economic inputs to outputs in organizations; specifically, this approach can explain how PROs use economic resources from government to support scientific outputs in turbulent markets.

---

[2] Cf. see for studies about technology, sources and effects of technology for industrial and economic change: Calabrese et al. 2002, 2005; Cavallo et al., 2014, 2014a, 2015; Coccia, 2002, 2004a, 2004b, 2005a, 2005c, 2005d, 2005e, 2005f, 2006, 2008c, 2009b, 2009c, 2010a, 2010b, 2010c, 2010d, 2010e, 2010f, 2011, 2012a, 2012b, 2012c, 2012d, 2012e, 2014, 2014b, 2014c, 2014d, 2014e, 2014f; 2015, 2015a, 2015b, 2015c, 2016, 2016a, 2017a, 2017b, 2017c, 2017d, 2017e, 2017f, 2017g, 2017h, 2017i, 2017l, 2017m, 2018, 2018b, 2018c, 2018d, 2018f, 2018g, Coccia et al., 2010, 2012; Coccia and Finardi, 2012; Coccia and Rolfo, 2002; Coccia and Wang, 2015, 2016.

[3] For other studies about organizational and managerial behaviour of public research labs, see: Calcatelli et al., 2003; Coccia, 2001, 2001a, 2003, 2004, 2005, 2005h, 2006a, 2008a, 2008b, 2009, 2009a, 2014a, Coccia and Bozeman, 2016; Coccia and Cadario, 2014; Coccia et al., 2015; Coccia and Rolfo, 2007, 2008, 2009, 2010, 2013.

The crux of the approach here is rooted in the metabolism and since this concept is uncommon in the studies of management, a brief background is useful to clarify it. In biology, the metabolism indicates the chemical processes that, in a living organism, transform food and drink into energy. The concept of metabolism has a vast use in several disciplines, such as industrial ecology, urban geography, ecological economics, economic geography, etc. (Kennedy *et al.,* 2007; Opschoor, 1997; Wolman, 1965). Marx (1867, 1978, 1981) considered the concept of socioecological metabolism as a circulation of financial, economic, human and natural resources to sustain economic systems through labor and capital. Nowadays, many scholars apply this concept to analyze the interrelations between socioeconomic and environmental factors (Rapoport, 2011). In economic geography, the concept of urban metabolism indicates the relations between economic and human resources of cities and environment to support their survival and/or grow over time (Fischer-Kowalski and Hüttler, 1999; Barles, 2010; Niza *et al.,* 2009).

This study here introduces the concept of economic metabolism of research organizations that can explain *how* these complex systems use and transform inputs (*e.g.,* research funds) into scientific outputs (publications, patents, innovative services, etc.). As a matter of fact, research organizations use economic and human resources to produce scientific research and knowledge (Coccia, 2001; 2012). The inputs of PROs are equipment, materials, research personnel, public and private funds, etc. Stephan (2010) argues that the increasing complexity of scientific research is generating a change of the capital-labor ratio within organizations for supporting science advances and discoveries. This change is also due to equipment and materials for research that are more and more costly. In particular, current scientific research labs have to invest in high-tech instruments for sustaining science and technology advances, such as Transmission Electron Microscopy that has supported the discovery of quasi-periodic crystals (Coccia, 2016). In the USA, academic institutions spent about

5 | P a g eCoccia M. (2018) How do public research labs use funding for research? A case study

*CocciaLab Working Paper 2018 – No. 32*

$2 billion in 2003 for research equipment, approximately 2.5 times the amount spent twenty years before (National Science Board, 2006). The importance of new materials, high-tech equipment and other inputs for research processes indicates that the access to substantial economic resources is a necessary condition for PROs to generate science advances, such as in medicine, nanotechnology, gravitational-wave astronomy, etc. (Coccia, 2014; cf., Stephan, 2010). For instance, National Science Foundation in the USA has done a huge investment of more than $1 billion for Laser Interferometer Gravitational-Wave Observatory (in construction, operational costs and research funds for scientists) for studying gravitational waves. In general, the cost of laboratories and scientific research is growing over time (cf., Stephan, 2012). Callon (1994) argues that state subsidy for public research labs is needed to investigate emerging research fields, though results can be uncertain and/or achieved only in the long run, such as measurement of gravitational waves and detection of their sources in the universe. One of the most important costs of research laboratories is human resources (Coccia and Rolfo, 2013; Coccia, 2014a). A small lab with about ten researchers in a U.S. university can have a cost of more than $500,000 to staff, before operational costs are added (cf., Stephan, 2012; Ehrenberg et al., 2003). Moreover, the managerial and organizational behavior of research labs is also affected by the regimes of public funding that are of two main types between countries (Billings et al., 2004): 1) institute oriented (or indirect funding of scientists). Many countries, in general, fund research institutes that in turn support scientific research of scientists, such as the CNRS in France and/or the CNR in Italy (Coccia, 2012); 2) scientific project oriented: direct funding of scholars in labs with a competitive grant system. This practice of funding scientists by projects is applied mainly in the United States and other Anglo-Saxon countries (*e.g.,* UK, South Africa, Australia, etc.).

In the USA, the largest contributor to research is the federal government with about 60 percent of funding, whereas local governments and industries contribute with about 6-7 percent (*cf.,* Bozeman and Crow, 1990). In the USA and Europe, during the 1990s and 2000s, the proportion of funds for research and education from government has increased, supporting new universities and/or expansion of older universities (OECD, 2017). In

6 | P a g eCoccia M. (2018) How do public research labs use funding for research? A case study

*CocciaLab Working Paper 2018 – No. 32*

Germany and Italy, Government-financed GERD (Gross domestic expenditure on R&D) as a percentage of Gross Domestic Product (GDP) has increased from 2000 to 2014, whereas in France and UK it has slightly reduced in the same period (OECD, 2017). In general, the Organization for Economic Co-operation and Development (OECD) countries have increased research funds from 0.61 in 2000 to 0.66 in 2014 (OECD, 2016). Italy has increased Government-financed GERD from 0.51 in 2009 to 0.54 in 2013. Moreover, levels of R&D personnel per thousand total employment in European countries have increased from 9.4 to 12.2 over 2000-2014 period. Italy also increased this level from 6.5 to 10.1 over the same period. Germany, France, UK and Spain have similar patterns of investment in human resources for research labs (OECD, 2016a). Overall then, research funding during 2000s has increased both in Europe and in the USA. However, economic crisis over recent years, especially in Europe, has led many governments to cope with huge government spending for research sector because of large infrastructures and vast structures of personnel created in the past (cf., Jacoby, 1994; Coccia and Rolfo, 2013). Current political economy of European countries is based on fiscal consolidation that affects a new public management in research sector based on downsizing of research institutes, their merger and a general shrinking of public research lab budgets (Coccia, 2012; Cruz-Castro and Sanz-Menéndez, 2016, 2018). The *leitmotiv* of this research policy is to reduce the general cost of public research bodies (cf., Coccia, 2012). These circumstances in current economic systems provide a valuable context for scholars of R&D management to analyze costs of public research organizations and investigate how public research funding is to produce research outputs[4]. This scientific topic is currently an under-researched field but it is basic for designing a strategy of R&D management directed to improve efficiency of public research labs and overall production of scientific research in the presence of scarce economic resources. This study here focuses on the biggest Italian research organization because it is a large public research body with an organizational behavior similar to other large European research bodies in Spain, France and Germany (Coccia, 2012). The investigation of this specific public research organization can clarify how public funding is spent by research labs in order to design best practices of R&D management directed to increase their efficiency over time. The methods to investigate these organizational and managerial issues are described in next section.

---

[4] For other factors affecting economic systems see: Chagpar and Coccia, 2012; Coccia, 2005b, 2005g, 2009d, 2010, 2013, 2013a, 2016b, 2017, 2017h, 2018a, 2018e, Coccia and Bellitto, 2018.

**Materials and Methods**

This study considers R&D organization as a complex system with (Coccia, 2001; Coccia et al., 2015; *cf.,* Brown and Svenson, 1998):

1. *Inputs:* human resources, equipment, funding for research, etc.
2. *Production process* transforms the inputs into outputs (economic metabolism).
3. *Outputs*: publications, software, patents, innovative outputs, etc.

The economic metabolism of research organizations is a process that describes how funding for research and other inputs are used and transformed in labs to produce research outputs

The system of R&D organizations is represented in Figure 1 that shows a linear model of economic metabolism of PROs.

Coccia M. (2018) How do public research labs use funding for research? A case study

*CocciaLab Working Paper 2018 – No. 32*

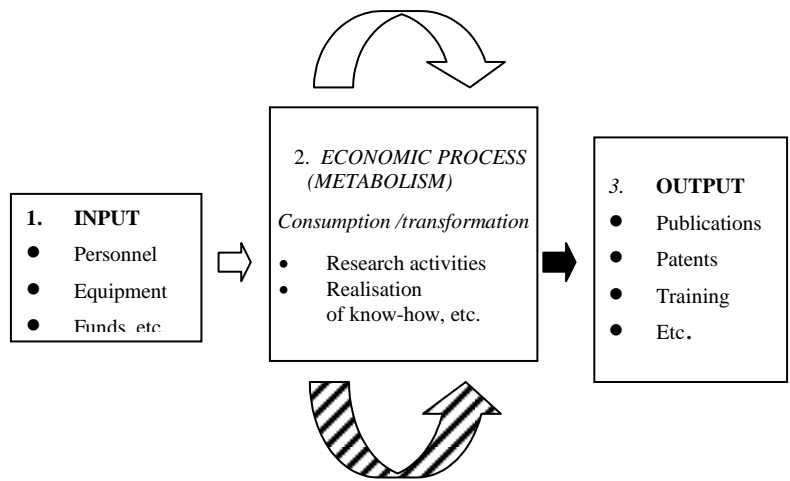

*Figure 1*. A linear model of the economic metabolism of public research organizations

This approach can identify how research funding is used within public research organizations to analyze possible causes of organizational inefficiencies. In particular, economic metabolism can help managers to analyze costs and design appropriate R&D management strategies directed to increase efficiency and productivity of research labs in the presence of scarce economic resources.

The approach of the economic metabolism of research organizations is applied here on the biggest public research body in Italy: The National Research Council of Italy or Consiglio Nazionale delle Ricerche (in short, CNR). The Italian CNR is an interesting case study because it is one of the largest European research bodies with an organizational behavior similar to other major European public research institutions, such as Le Centre National de la Recherche Scientifique –CNRS-in France and Consejo Superior de Investigaciones Científicas – CSIC-in Spain, etc. The study here can provide interesting insights to understand how public resources of research bodies are used to support R&D management practices to cope with shrinking public research lab budgets in turbulent markets. In 2018 the organizational structure of the CNR has seven departments:

- Earth system science and environmental technologies
- Biology, agriculture and food sciences
- Chemical sciences and materials technology
- Physical sciences and technologies of matter
- Biomedical sciences
- Engineering, ICT and technologies for energy and transportation
- Social sciences and humanities, cultural heritage

These seven departments include about 120 research institutes and several decentralized units. The CNR in 2015 had about 8,372 total units of personnel that were organized in these institutes operating in a broad range of scientific fields and technological applications (Coccia, 2004, 2005, 2008a, 2008b, 2012; Coccia and Rolfo, 2007; cf., Central Management of Human Resource of the CNR, 2017). Current organization of the CNR is the results of many reforms started in 2001 and ongoing. Common features of these reorganizations, based on a new public management approach, are the merger of several institutes and research units for reducing general costs (Coccia, 2012). These reforms are also introduced with the hope to increase the market orientation and efficiency of this public research organization. However, one of the major consequences of these reforms is an internal financial crisis since public funds are no longer sufficient to cover current expenses of institutes. In addition, the restructuring has increases bureaucratization and coordination costs within and between institutes and decentralized research units (Coccia, 2012; Coccia, 2009, 2009a).

## 1.1 Methodological steps for acquiring and analyzing the utilization of public funds

The sources of data are the annual financial statements of the CNR (Consiglio Nazionale delle Ricerche, 2016). The methodological steps for acquiring data are:

1. Download of financial statements (income statements, balance sheet and cash flow statement) from the official homepage of the National Research Council of Italy (Consiglio Nazionale delle Ricerche-CNR) for the period 1997-2016 (see, https://www.cnr.it/). These documents are in PDF format.

2. In particular, this study focuses on the income statement that is one of the fundamental financial statements. The income statement shows revenues, expenses, and profit/loss of the fiscal year for this public research organization (PRO).

3. Items acquired from the income statement of CNR in Italy are in table 1.

**Table 1.** Items from the income statement of the CNR (1997-2016)

| *Description* |
| --- |
| • Year/Euros |
| • Total Revenue (State Subsidy and Contracts) |
| • Cost of Personnel |
| • Other Costs |
| • Total Cost |
| • Materials and Products |
| • Service |
| • Leased Assets of Third Parties |
| • Salary |
| • Social Security Taxes |
| • Severance Pay |
| • Other Costs |
| • **Economic Surplus/Loss of Fiscal Year** |

4. From 1997 to 2001 data are in Italian currency Lire, from 2002 to 2015 data are in currency Euro. In order to create a comparable framework, data of the income statement from 1997 to 2001 in Lire currency are converted in Euro currency with the rate of change fixed by European Central Bank: 1Euro=1936.27 Lire

5. Data of items in table 1 are reported in a spreadsheet Excel, where first line has years from 1997 to 2015 and first column has the list of items indicated in table 1.

Coccia M. (2018) How do public research labs use funding for research? A case study

*CocciaLab Working Paper 2018 – No. 32*

6. Structure of data in the spreadsheet of Excel is available to download, reuse, reproduce analyses and/or view at Coccia (2018).

*1.2  Data analysis and procedures*

The preliminary statistical analysis of the economic metabolism of PROs is given by descriptive statistics, trends and bar graphs that represent arithmetic mean (or number of cases) of variables on *y*-axis and some inputs on *x*-axis.

The main statistical analysis of economic metabolism of PROs is a set of techniques given by:

- A linear model of simple regression analyses trends of costs/revenue of PRO. The specification of the model is:

$$Y_t = \lambda_0 + \lambda_1 x_t + u_t \qquad t = time \qquad [1]$$

  where:
  - $Y_t$ = total revenue or cost of personnel (dependent variables)
  - $x_t$ = time
  - $u_t$ = error term

These models are estimated with the Ordinary Least Squares (OLS) method. The goodness of fit is measured with the coefficient of determination $R^2$.

- The following index M (in short, Metabolism) explains the dynamics of the economic metabolism of public research labs based on specific costs of income statement, such as cost of personnel:

$$M_t = \left( \frac{cost\ of\ personnel}{Total\ revenue} \% \right)_t \qquad t = time \qquad [2]$$

In particular, Eq. [2] indicates the proportion of cost of personnel on total revenue, a main indicator of the consumption of economic resources (public funds) in PROs over time.

- Another index for analyzing the economic metabolism here is the rate of arithmetic growth of total revenue (state subsidy and public contracts), cost of personnel and total cost. In particular, if the level of these variables at beginning is $_aP$ (1997) and at the end of the period is $_tP$ (2015), and the time interval is equal to $t$ (2015-1997=19), the rate of arithmetic growth $^ar$ of variables under study is given by:

$_tP =\, _0P +\, _0P(^ar \cdot t)$ where $_tP -\, _0P =\, _0P\ ^ar \cdot t$ and hence

$$^ar = \frac{_tP -\, _0P}{_0P \cdot t} = \text{rate of arithmetic growth} \qquad [3]$$

- Moreover, the economic metabolism also applies an allometric model of growth in order to analyze the growth of costs on total revenue (public funds). The model is applied here on cost of personnel as follow:

Let $X(t)$ be the extent of the cost of personnel of research organization $i$ at the time $t$ and $Y(t)$ be the extent of the total revenue of research organization $i$ at the same time. If both $X$ and $Y$ increase according to some S-shaped pattern of growth, such a pattern analytically can be represented with the differential equation of logistic function:

$$\frac{1}{Y}\frac{dY}{dt} = \frac{b_1}{K_1}(K_1 - Y)$$

This equation can be rewritten with the following simple model of growth (*see*, Coccia and Bozeman, 2016):

$$X = A_1(Y)^B$$

The logarithmic form of this equation $X = A_1(Y)^B$ is a simple linear relationship:

$$LnX = LnA_1 + BLnY$$

The specification of the model in this study is:

$Ln\ x_t = Ln\ a + B\ Ln\ y_t + u_t$ \quad (with $u_t$ = error term) \qquad [4]

where:

*a* is a constant

$x_t$ will be the extent of the cost of personnel of research organization *i* at time *t*

$y_t$ will be the extent of the total revenue of research organization *i* at time *t;* it is a driving force of the cost of personnel over time.

The coefficient *B* of the Eq. [4] indicates:

− *B* = 1, both cost of personnel and total revenue of research organization are growing at the same rate (*isometric growth*). This indicates a normal economic metabolism of research labs.
− *B* < 1, the rate of cost of personnel is growing more slowly than that of total revenue: negative allometric growth. This indicates an efficient economic metabolism of research labs
− *B* > 1, there is a disproportionate growth of the cost of personnel over time. This coefficient indicates a pathological economic metabolism of research labs.

The analysis of the economic metabolism, just described, can detect the dynamics of costs in research organizations to support best practices of R&D management for increasing scientific production and efficiency. Statistical analyses are performed by using the Statistics Software SPSS® version 15.0

**Results**

Results of this study are based on financial and economic data from the income statement (1997-2015 period) of the Italian National Research Council (CNR), one of the biggest European public research organization (Coccia, 2018). Figure 1 shows main components of costs of the CNR and the highest level of consumption is given by the cost of personnel, followed by service, materials and products (cf., Figure 1A in Appendix). In particular, the cost of personnel is driven mainly by salary and social security taxes (cf., Figure 2A in Appendix). As a consequence, the analysis here focuses on critical factor of the cost of personnel to assess processes of economic metabolism within this public research organization.

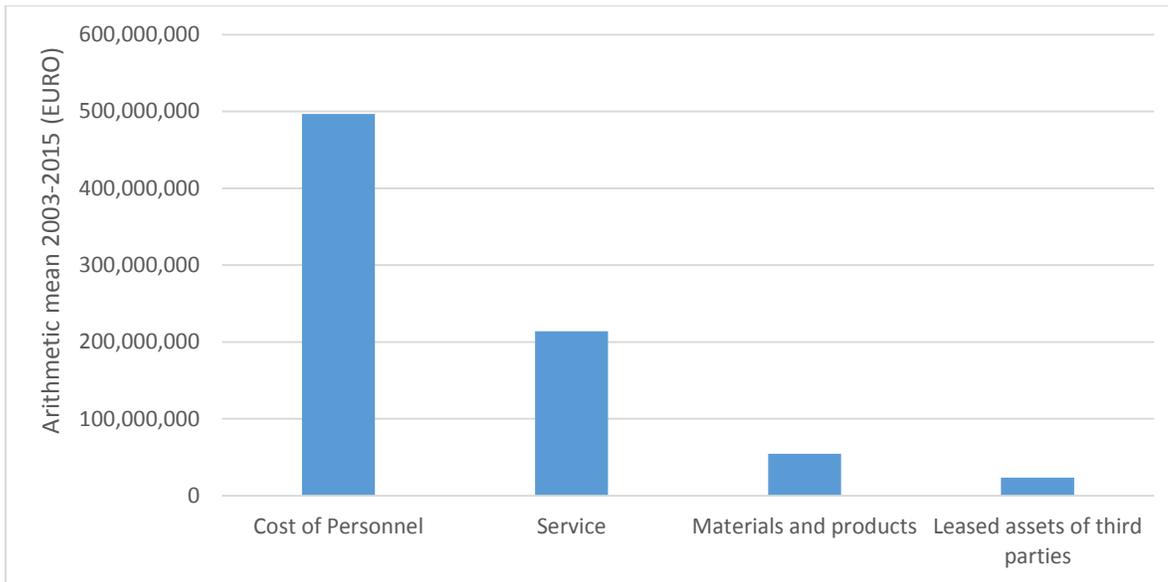

*Figure* 1. Economic metabolism of CNR organization:
Arithmetic mean of main costs over time based on CNR data (Coccia, 2018)

In particular, Figure 2 shows that the trend of the cost of personnel has a higher growth rate than total revenue. Table 2 shows results of the regression analysis based on equation [1]: the standardized coefficient of regression for the cost of personnel is 0.93 *vs.* 0.83 of total revenue. Moreover, the standardized coefficient of regression of the total cost is slightly higher than total revenue: 0.85 *vs.* 0.83. (cf. also, Table 2 and Figure 3). Figure 3A in Appendix shows the trend of other costs.

Coccia M. (2018) How do public research labs use funding for research? A case study

*CocciaLab Working Paper 2018 – No. 32*

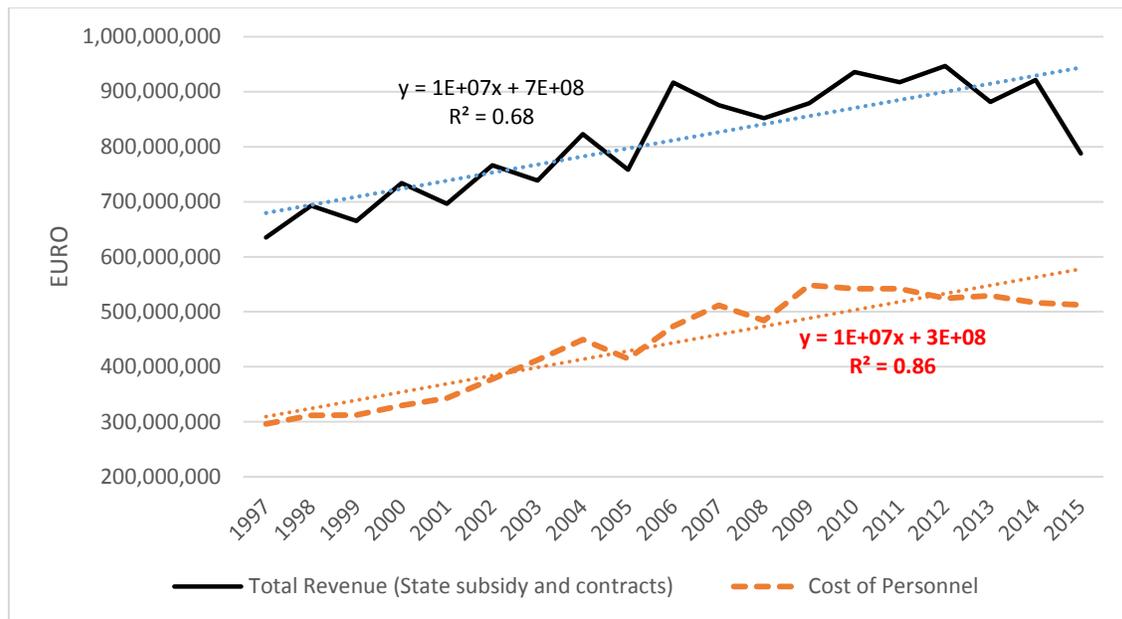

*Figure* 2. Economic metabolism of CNR organization: Total revenue and cost of personnel over time based on CNR data (Coccia, 2018)

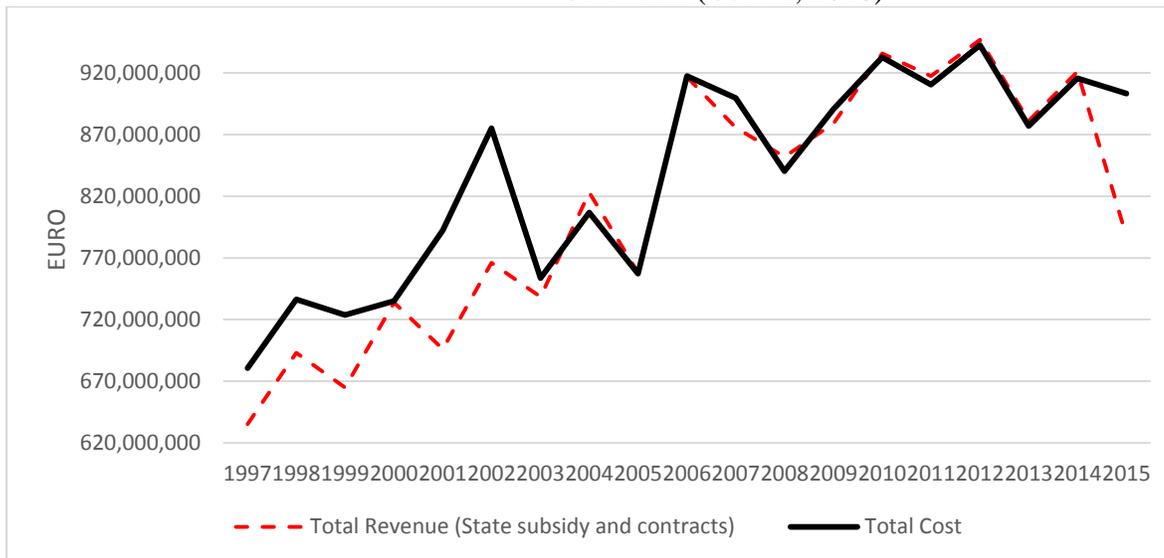

*Figure* 3. Economic metabolism of CNR organization: Comparison of total revenue and total cost over time based on CNR data (Coccia, 2018)

Coccia M. (2018) How do public research labs use funding for research? A case study

*CocciaLab Working Paper 2018 – No. 32*

Table 2: Estimated relationships of the economic metabolism of CNR organization

|  | Constant $\lambda_0$ (St. Err.) | Coefficient $\lambda_1$ (St. Err.) | Stand. coefficient | $R^2$ | F (Sign.) |
|---|---|---|---|---|---|
| 1. Dependent variable (D): Total Revenue (State subsidy and public contracts) | | | | | |
| Explanatory variable<br>○ Time 1995-2015 | −28637583293.247*** (4877200736.575) | 14680572.412*** (2431297.386) | 0.83 | 0.66 | 36.46 (0.001) |
| 2. Cost of Personnel (D) | | | | | |
| Explanatory variable<br>○ Time 1995-2015 | −29461869415.987*** (2904961812.461) | 14908079.764*** (1448131.099) | 0.93 | 0.84 | 105.98 (0.001) |
| 3. Total Cost (D) | | | | | |
| Explanatory variable<br>○ Time 1995-2015 | −24481539439.878*** (3854266454.615) | 12621066.402*** (1921361.958) | 0.85 | 0.70 | 43.15 (0.001) |

*Note*: ***=Coefficient is significant at *p-value<0.001*

In particular, from 2002, after the European Monetary Unification in 2001, the cost of personnel of CNR organization has sharply increased. In fact, Figure 4 shows that the proportion M of the cost of personnel on total revenue (eq. [2]) is increasing over time; this analysis of economic metabolism reveals the pathologic growth of the cost of personnel on total revenue that in 2015 absorbs more than 65% of total revenue. Figure 4 also confirms the critical point of the year 2002 with an inversion of trends between cost of personnel and other costs, and as a consequence the problematic economic metabolism of this public research organization that absorbs a high share of public research funding for sustaining the structure of personnel over time.

Table 3 shows the growth rate based on eq. [3] that it is very high for the cost of personnel and high for the total cost of CNR in comparison with total revenue (from 1997 to 2015 period).

**Table 3.** Economic metabolism of CNR organization: rate of arithmetic growth of specific items from annual income statements (for data see Coccia, 2018)

| Rate of arithmetic growth | Total revenue | Cost of personnel | Total cost |
|---|---|---|---|
| [a]$r$ (1997-2015) | 118.72 | 167.87 | 127.44 |

Coccia M. (2018) How do public research labs use funding for research? A case study

*CocciaLab Working Paper 2018 – No. 32*

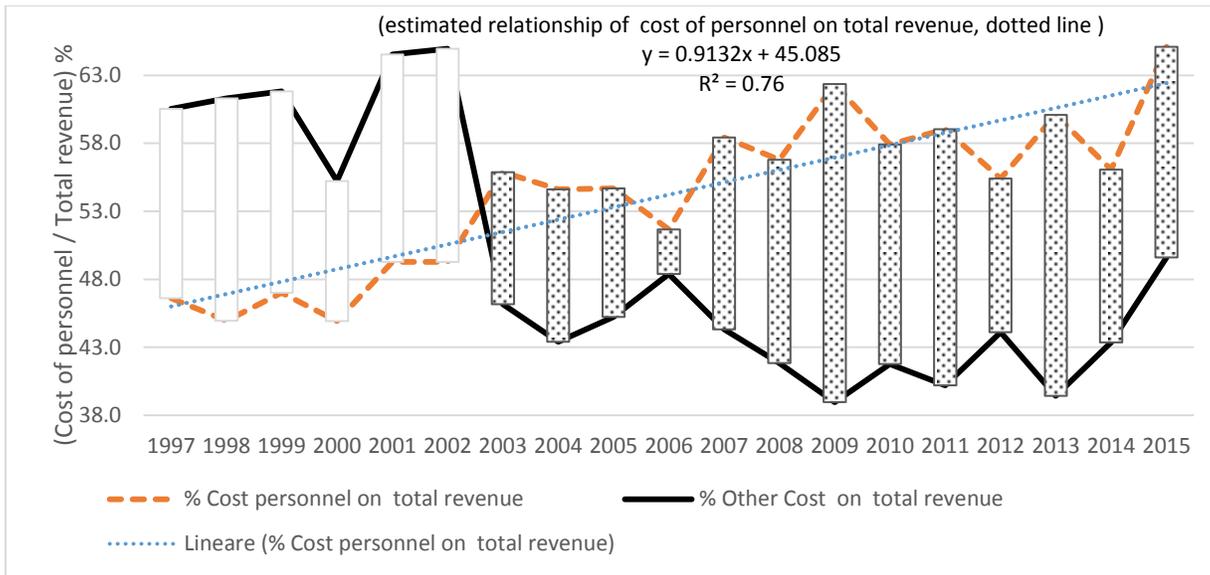

*Figure* 4. Economic metabolism of CNR organization: proportion of the cost of personnel on total revenue over time based on CNR data (Coccia, 2018)

Results of the allometric model [4] in table 4 show that the cost of personnel has B >1. This finding indicates a disproportionate growth of the cost of personnel on total revenue within the CNR organization and confirms a pathologic economic metabolism of this research organization under study. This result is consistent with other analyses.

**Table 4**: Economic metabolism of CNR organization: Estimated relationship of equation [4]

| | *Constant* $\lambda_0$ *(St. Err.)* | *Coefficient* $B=\lambda_1$ *(St. Err.)* | *Stand. coefficient* | $R^2$ *(Std. error of the Estimate)* | *F (Sign.)* |
|---|---|---|---|---|---|
| Dependent variable: LN Cost of personnel | | | | | |
| *Explanatory variable* LN Total Revenue (State subsidy and public contracts) | −13.04** (3.14) | 1.61*** (0.15) | 0.93 | 0.86 (0.082) | 110.14 (0.001) |

*Note*: ***=Coefficient is significant at *p-value<0.001*

To sum up, statistical evidence reveals that the economic metabolism of this public research organization absorbs a substantial proportion of public funds (state subsidy and public contracts) to cover the high cost of personnel over time. Overall, then, the approach of economic metabolism applied on this research institution seems to show possible organizational inefficiencies, driven by high cost of personnel.

**Discussion and concluding observations**

The economic metabolism of research organization under study here reveals some critical organizational issues. In response to the first question stated in the introduction, it seems that the economic metabolism of this public research organization absorbs a high share of the total revenue for the cost of personnel. This result can be due to a moderate growth of research personnel in the presence of public budget cuts over time.

With respect to the second question, the answer also seems clear: consumption of economic resources for the cost of personnel is growing over time. To reiterate, the critical factor is the high cost of personnel and R&D management has to control (feedback mechanisms) this factor in the presence of shrinking public research lab budgets to support efficiency and sustainability of research labs in the long run.

In particular, the results of this analysis suggest:

1. A growing trend of the cost of personnel from 1997 onwards. In particular, the proportion of the cost of personnel on total revenue (state subsidy) has attained an very high level of more than 65 % in 2015 (based on CNR data).

2. The rates of arithmetic growth over 1997-2015 reveal that total revenue (state subsidy and public contracts) has grown by 118.72%, whereas the cost of personnel by 167.88% and total cost by 127.44%. The imbalance of growth between cost of personnel and total revenue within this organizational body is causing organizational inefficiencies for scientific production.

3. Before the European Monetary Unification[5], Italian CNR had the proportion of the cost of personnel on total revenue lower than other costs. From 2002 onwards, there is an inversion of tendency with a growing incidence of the cost of personnel on total revenue: in 2015, the 65 percent of the total revenue (state subsidy) is absorbed by the cost of personnel, whereas in 1997 –before the European Monetary Unification- this share was about 47%.

---

[5] Cf., Coccia, 2017i for some economic effects of European Monetary Unification.

Coccia M. (2018) How do public research labs use funding for research? A case study

*CocciaLab Working Paper 2018 – No. 32*

In short, the economic metabolism of this public research organization reveals a problematic consumption of public funds for the cost of personnel that can be the source of several inefficiencies. The increasing share of the cost of personnel of this public research organization on total revenue (state subsidy) can be due to exogenous and endogenous factors.

The exogenous factors can be the socioeconomic change in Europe with the European Monetary Unification that has generated a shock of prices and costs, with effects on organizations and consumers (cf., Coccia, 2016; 2017). One of the principal consequences on research sector is the cuts in public spending because of high government debt of Italy (Coccia, 2013, 2016, 2017). In particular, shrinking public research lab budgets of CNR have reduced the ability of this research organization to adapt to new situations and increased the levels of uncertainty within the national and European system of innovation.

The endogenous factors can be due to the increasing bureaucratization and coordination costs generated by restructuring this large public research body from 2001 and ongoing (Coccia, 2009, 2009a, 2012; *cf.,* Bozeman, 2000).

The difficult organizational behavior of this public research organization is also due to current funding regime of public research in Italy. In particular, government spending for public research bodies in Italy, such as CNR, supports mainly the salary of researchers (and other personnel) and at small extent other expenses of institutes. Moreover, in Italy, the funds for projects based on a competitive grant system are scarce to support the scientific activities of all CNR researchers. As a consequence, funding system of paying only the salary of research personnel is not appropriate to support modern scientific research that is more and more based on expensive materials and high-tech equipment for supporting research processes (Coccia, 2014a; Stephan, 2010).

The Italian context also presents the high pressure of labor unions that, in the presence of shrinking public research lab budgets, continue to support hiring of personnel and increases of salaries without considering the necessity of funding the research processes of scientists (Coccia and Rolfo, 2009, 2010; *cf.,* Coccia and Cadario, 2014). This policy may be one of the contributing factors that reduces productive investments for scientific production and career advancements of current personnel to improve their job satisfaction and productivity (*cf.,* Coccia, 2001a; Bozeman and Gaughan, 2011; Bozeman *et al.,* 2001; *cf.,* Rainey and Bozeman, 2000). Hence, the approach of economic metabolism shows organizational deficiencies of CNR because of lack of appropriate economic resources for scientific research and increases of the cost of personnel on total revenue. In short, this result indicates that public research system in Italy has a rigid organization to cope with economic crises and adapt to a turbulent environment with shrinking public research lab budgets.

The economic metabolism, presented here, is critical to cost analysis and management within public research organizations. This approach can support R&D management that, in the presence of budget cuts, can improve the allocation of resources and the efficiency of research organization by controlling costs, and balancing the funding for research and for human resources development. Moreover, if public research bodies cannot offer substantial extrinsic incentives to scientists based on high pay similar to private sector because of scarce public funds, then R&D management should increase intrinsic rewards to support motivation of researchers and their scientific performances (Belle and Cantarelli, 2015; Weibel et al., 2010; cf., Ryan and Deci, 2000, 2000a; Keller, 2017; cf., Benati and Coccia, 2017, 2018; Coccia and Benati, 2017, 2018). O'Reilly et al. (1991) have suggested that intrinsic rewards may support affective commitment, job involvement and satisfaction of subjects in organizations. Specifically, intrinsic rewards may support productivity of research personnel also in the presence of budget cuts (Frank and Lewis 2004; Wright , 2007, p. 60). In this context, Crewson (1997, pp. 503–4) argues that: "Intrinsic rewards are more important to public employees than to those employed in the private sector".

To conclude, the present study here is exploratory in nature and findings need to be considered in light of their limitations. This study has focused on a main input within PROs (state subsidy and public contracts) but a

Coccia M. (2018) How do public research labs use funding for research? A case study

*CocciaLab Working Paper 2018 – No. 32*

comprehensive analysis of the economic metabolism of research organizations should also consider in future other inputs and outputs. The results of this study should also be compared in future to other research organizations in Europe, such as CNRS in France and CSIC in Spain in order to detect similar and/or different organizational behavior for appropriate research policies (*cf.,* Rainey and Bozeman, 2000; Crow and Bozeman, 1998). This study suggests that public research organization under study (i.e., CNR) absorbs a high share of the total revenue (state subsidy) for the high cost of personnel generating organizational deficiencies and future problems of organizational sustainability. However, the results of this study are of course tentative, since we know that other things are often not equal over time and space. Overall, then, this paper focuses on a specific case study clearly important but not sufficient for broader understanding and generalization of organizational and managerial behavior of public research labs in turbulent markets with scarce economic resources. Hence, there is need for much more detailed research in future, using data within and between countries, to shed further theoretical and empirical light on consumption of public research organizations in the presence of shrinking public research budgets.

**Appendix**

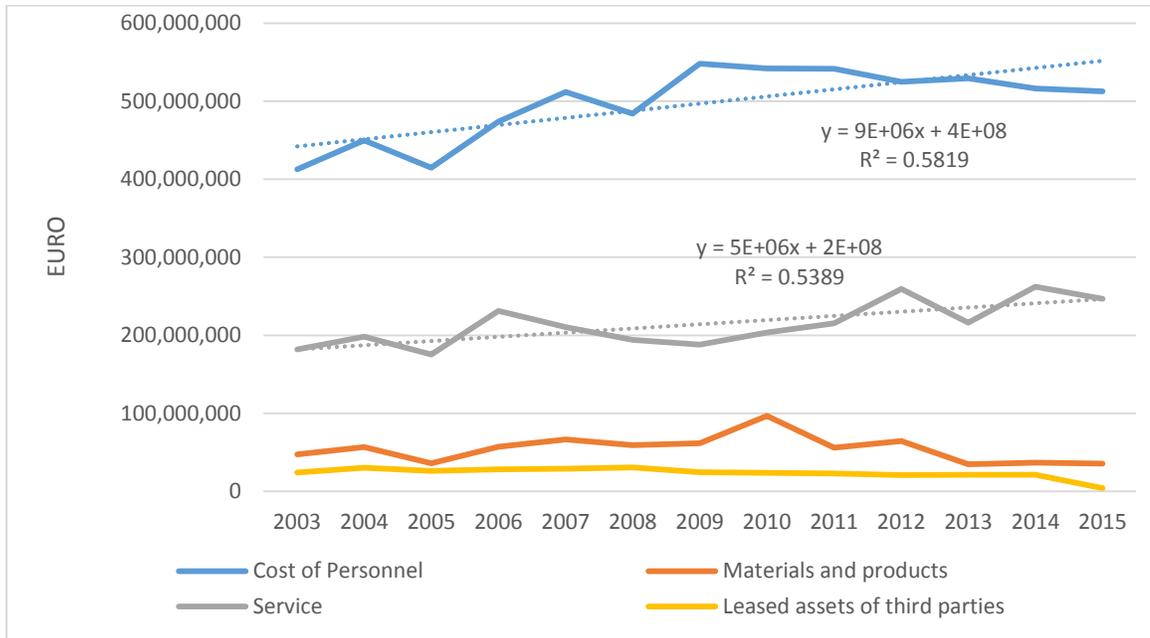

*Figure* 1A. Trend of main costs over time based on CNR data (Coccia, 2018)

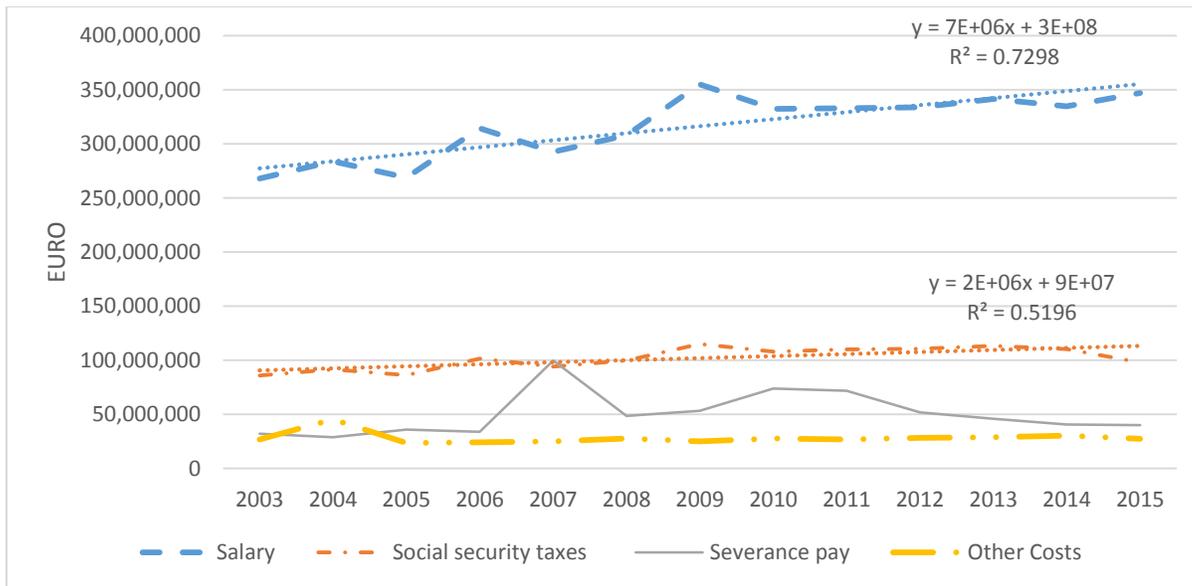

*Figure* 2A. Trend of main elements of the cost of personnel over time based on CNR data (Coccia, 2018)

Coccia M. (2018) How do public research labs use funding for research? A case study

*CocciaLab Working Paper 2018 – No. 32*

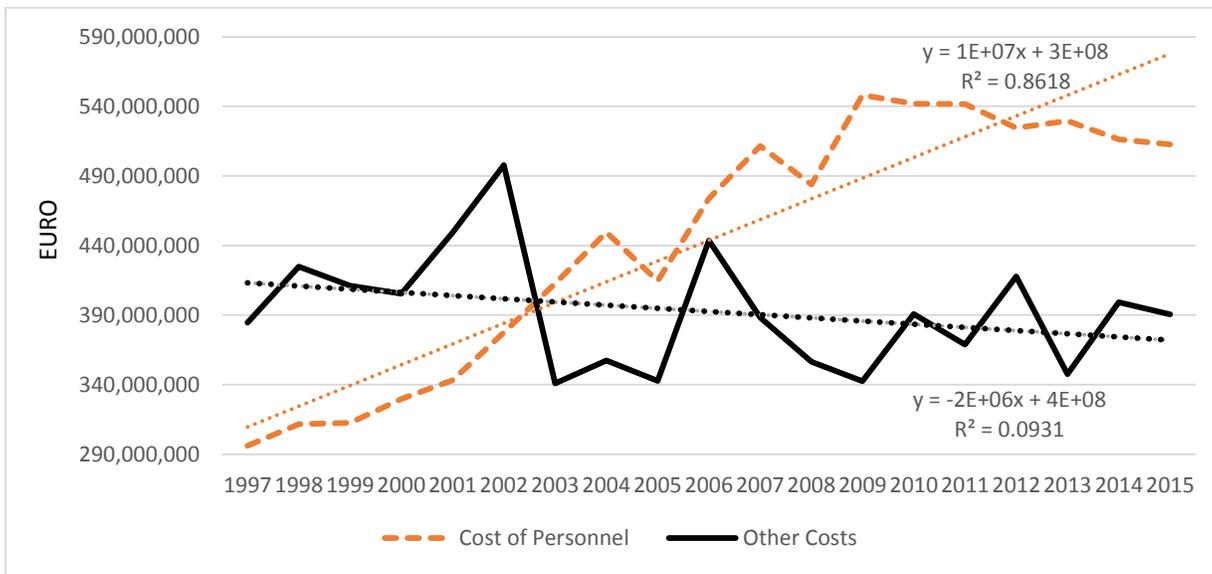

*Figure* 3A. Cost of personnel and other cost over time based on CNR data (Coccia, 2018)

Coccia M. (2018) How do public research labs use funding for research? A case study

*CocciaLab Working Paper 2018 – No. 32*

Coccia M. (2018) How do public research labs use funding for research? A case study

*CocciaLab Working Paper 2018 – No. 32*